\documentstyle [12pt]{article}
\newcommand{\eqn}{\begin{equation}}
\newcommand{\eqnend}{\end{equation}}

\begin{document}
\baselineskip 0.8truecm

\centerline{\bf NEGATIVE 4-PROBE CONDUCTANCES OF MESOSCOPIC,}
\centerline{\bf SUPERCONDUCTING
WIRES.}
\vskip 1truecm
\centerline{\bf SJ Robinson\dag,  CJ Lambert\dag and M Jeffery\ddag}
\vspace{0.5cm}
\centerline{\bf \dag School of Physics and Materials}
\centerline{\bf University of Lancaster, Lancaster, LA1 4YB, UK}
\vspace{0.5cm}
\centerline{\bf \ddag Goto Laboratory, The Institute of Physical and Chemical}
\centerline{\bf Research (RIKEN), Hirosawa 2-1, Wako-shi, Saitama 351-01,
Japan}

\vskip 1truecm
{\bf ABSTRACT}

We analyze the longitudinal 4-probe
conductance of mesoscopic normal and
superconducting wires and predict that in the superconducting case,
 large negative values
can arise for both the weakly disordered and localized regimes.
This contrasts sharply
with the behaviour of the longitudinal 4-probe conductance of normal wires,
which in the localized limit is always exponentially small and positive.


\pagebreak

In contrast with the huge literature on
quantum transport in normal, phase coherent,
structures~\cite{review1,review2}, a detailed knowledge of
mesoscopic superconductoring structures is only now being developed.
Such structures constitute new quantum objects, which during
the past 3 years, have yielded
many surprises~\cite{petrashov1,petrashov2,beenakker}. Most theoretical work to
date has focussed on
two-probe transport coefficients, such as
normal-superconducting (N-S) boundary~\cite{btk}
conductances and N-S-N 2-probe conductances~\cite{here:2probe91,here:2probe93}.
These are simpler to analyze than multi-probe coefficients and therefore
without good reason for expecting new physics to emerge, there would seem to be
little point in injecting unnecessary details into an already
complex theory.
The aim of this Letter is to demonstrate that multi-probe conductances
contain new features which are absent from  two-probe measurements.
A key result of our work is that when a normal system with a
positive 4-probe conductance is allowed to become superconducting,
the sign of the conductance can change.

As an example, we study the 4-probe conductances $G_{jk,lm}$ of the
2 dimensional wire shown in figure 1a, where
$G_{jk,lm}=I_{jk}/(V_l-V_m)$, with $V_l-V_m$ the potential difference
between voltage probes $l$, $m$ and $I_{jk}$ the current flowing from probe
$j$ to probe $k$. We consider the case for which the system size is
smaller than the inelastic scattering length and
focus attention on the longitudinal conductances
$G_{13,24}$ and $G_{14,23}$.
It has been shown~\cite{buttiker:multi88} that for normal
materials, both negative and positive multiprobe
conductances can both occur depending on the geometry of the device.
On the other hand, numerical simulations of electron waveguide
couplers~\cite{wang:pc92},
have found that for clean materials, the analogues of
 $G_{13,24}$ and $G_{14,23}$  are always positive, in
agreement with theoretical arguments valid for devices satisfying certain
spatial symmetries~\cite{wang:pc92,avishai:res89}.

In this letter we first derive a general
 criterion which determines the sign of the 4-probe
conductance in {\em any} arrangement of normal conductors.
In the case of
localized normal wires, for which the quasi-particle transmission probability
from
one end of the sample to another is exponantially small, this shows that
  $G_{13,24}$ and $G_{14,23}$ are
both positive
and of order $\sim \exp -2L/\xi$, where $\xi$ is the localization length.
In contrast, using a recently derived generalisation~\cite{here:multi93}
of the multiprobe B\"uttiker formulae to the case in which Andreev
scattering is permitted, we show that for a
{\em superconducting} wire with exponentially small quasi-particle
transmission,
longitudinal conductances are finite
and may be of either sign. We also present numerical
results for the behaviour of 4-probe conductances
in the presence of finite quasi-particle transmission along the wire.

Consider first the case of a normal wire, for which
 it is known~\cite{buttiker:multi88} that
\eqn G_{jk,lm} = D/(T_{jl}T_{km}-T_{jm}T_{kl}) \label{eq:G}\eqnend
where $T_{ij} = \sum_{\alpha\beta}T_{ij}^{\alpha\beta}$ and
$T_{ij}^{\alpha\beta}$ is the probability for a particle incident in
channel $\beta$ of probe $j$ to be transmitted to  channel $\alpha$ of
probe $i$.
$D$ is the determinant of the matrix obtained by crossing out any
one row and column of the matrix of transport coefficients
$A$, $A_{ij}=N\delta{ij} - T_{ij}$ where $N$ is the number of channels
in each probe (assumed equal). The value of this determinant is
independent of which row and column are removed~\cite{buttiker:rev92},
although because of the relation $\sum_i A_{ij} = \sum_j A_{ij} = 0$
the expression may be written in many equivalent ways.
 In order to make the analysis clearer
we will write those scattering coefficients involving
transmission along the wire, for example $T_{13}$, $T_{24}$
(but not eg. $T_{12}$ or $T_{34}$) in the form
$T_{ij}= t_{ij}\exp{-2L/\xi}$ where $t_{ij}$ has magnitude of order
unity.  We expand $D$ by removing row 4 and column 4 of $A$ and
substituting $A_{ii} = -\sum_{j \neq i} A_{ij}$. This yields
\begin{eqnarray}
 D  & = & \exp -2L/\xi \; (T_{12}t_{23}T_{34} + T_{12}t_{24}T_{34}
+ t_{13}T_{21}T_{34} + t_{14}T_{21}T_{34}) \nonumber\\
 & & + \exp -4L/\xi \; (T_{12}t_{24}t_{31} + T_{12}t_{24}t_{32} +
 T_{34}t_{13}t_{23} + T_{34}t_{13}t_{24} \nonumber\\
 & & \;\;\;\;\;\;\;\;\;\;\;\;\;\;\; +T_{21}t_{14}t_{31} + T_{21}t_{14}t_{32} +
 T_{34}t_{14}t_{23} + T_{34}t_{14}t_{24} ) \nonumber\\
 & & + \exp -6L/\xi \; (t_{13}t_{24}t_{32} + t_{14}t_{23}t_{31} +
 t_{14}t_{24}t_{31} + t_{14}t_{24}t_{32})\;\;\; , \label{eq:D}\end{eqnarray}
which demonstrates that for $L >> \xi$, $D$ decays exponentially with $L$.
Although we have written $D$ explicitly as a sum of powers of
$\exp -2L/\xi$, expression~(\ref{eq:D}) is exact and remains true when
$L \ll \xi$.
The key observation here is that $D$ is the sum of positive terms
and hence is {\em always} positive. Hence the sign of $G_{jk,lm}$
depends only on the relative magnitudes of $T_{jl}T_{km}$ and $T_{jm}T_{kl}$.
This was noted separately by Avishai and Band~\cite{avishai:res89}
 for a crossed wire
arrangement and by Wang {\it et al}~\cite{wang:pc92} for ballistically coupled
wires.
However both of these references rely on being able to apply certain
symmetries to the system, which simplifies the form of $D$. The above
analysis shows that $D$ is positive, independent of such symmetries.

Now consider the longitudinal conductances $G_{13,24}$ and $G_{14,23}$
in the limit $L \gg \xi$.
Since none of $T_{12}$, $T_{34}$ and $T_{43}$ are expected to decay
with $L$, the denominator of expression~(\ref{eq:G}) for these conductances
is positive, with magnitude of order unity. Hence  in this limit
\eqn G_{13,24} \sim G_{14,23} \sim  + \exp -2L/\xi \eqnend
Since the two-probe conductance $G_2$ of a device with substantial
localization is of order $\sim \exp -2L/\xi$, we see that for
normal wires in the localised limit 4-probe conductance measurements
will give results of the same sign and order of magnitude as 2-probe
measurements.

Conductance
formulae of the kind shown in equation (1), describe a normal scatterer
connected to normal probes. The generalisation of this
approach to the case where the scatterer incorporates
superconductivity, but the probes remain normal,
was first derived for 2 probes
by ~\cite{here:2probe91,here:2probe93} and more
recently for many probes by Lambert, Hui and
Robinson~\cite{here:multi93}. This generalisation leads to the introduction of
a matrix of transport coefficients
$A$, which at zero temperature, has elements
 $A_{ij} = N\delta_{ij} -T^O_{ij} + T^A_{ij}$,
where superscripts $O$ and
$A$ refer respectively to normal and Andreev scattering.
It is shown in~\cite{here:multi93} that in general
\eqn G_{ij,kl} = d/(b_{ik}-b_{jk}- b_{il} + b_{kl}) \eqnend
Here, $d = \det A$, and $b_{mn}$ is the cofactor of the matrix element
$A_{mn}$.
In order to highlight the
behaviour of $G$ in the localized limit we work to zero'th order
in $L/\xi$ so that $A$ becomes
\eqn A = \left( \begin{array}{cccc}
    N - T^O_{11} + T^A_{11} & -T^O_{12} + T^A_{12} & 0 & 0 \\
     - T^O_{21} + T^A_{21} & N-T^O_{22} + T^A_{22} & 0 & 0 \\
    0 & 0  & N - T^O_{33} + T^A_{33} & -T^O_{34} + T^A_{34}  \\
    0 & 0  &  -T^O_{43} + T^A_{43} & N-T^O_{44} + T^A_{44} \end{array}
    \right) \eqnend
The respective determinants $d_{TL}$ and $d_{BR}$ of
the top left and bottom right blocks of $A$  are both positive, because
in general $N= \sum_{j=1}^4 (T^O_{ij}+T^A_{ij})$ for any $i$,
so that $A_{11}>A_{12}$, $A_{22}>A_{21}$ etc. Hence $d=d_{TL}d_{BR}$ is
positive
in the localised limit.

Now consider the longitudinal conductance
$G_{13,24} = d/(b_{12}+b_{34}- b_{13} - b_{24})$. To zeroth order
in $L/\xi$ we have $b_{13}=b_{24}=0$, $b_{12}=-a_{21}d_{BR}$, and
$b_{34}=-a_{43}d_{TL}$. Hence
\eqn G_{13,24} = -\left( \frac{T^A_{21}-T^O_{21}}{d_{TL}}
+ \frac{T^A_{43}-T^O_{43}}{d_{BR}} \right)^{-1}  \label{eq:scG}\eqnend
A similar expression can be derived for $G_{14,23}$
Since all the terms on the right hand side of~(\ref{eq:scG})
 are of order unity,
$G_{13,24}$ would also be expected to be of this order. However, in contrast
with the behaviour of normal systems the conductance
can be of arbitrary sign, depending on the relative magnitudes of
the normal and Andreev scattering coefficients for transmission from
a current to a voltage probe at the same end of the wire. In general
stronger Andreev scattering favours negative longitudinal conductances,
while stronger normal scattering favours positive conductances. It
should also be noted that it is possible in principle for the two terms
in~(\ref{eq:scG}) to approximately cancel, so that the
longitudinal conductance may become arbitrarily large, {\em even in
the strongly localised limit}~\cite{note}.

To test whether
finite negative multi-probe conductances can occur in practice
we have performed numerical simulations of both
normal and superconducting multiprobe structures.
 Following Avishai and
Pichard {\it et al}~\cite{avishai:graphs93,pichard:graphs93}, we model a 2d
wire with 3 channels and 4 probes by a network of 1d wires connected at nodes,
shown in figure 1b. In the figure $M$ denotes the number of slices between
the pairs of probes. The figure shows the case $M=10$, but simulations were
performed over the range of lengths $M=1$ to $M=40$ for the normal wire and
$M=1$ to $M=100$ for the superconducting wire.
Disorder is introduced by specifying random scattering matrices
at each node, while allowing perfect transmission along each 1d wire.
For a normal system, the $N_w\times N_w$ scattering
matrix of a node connecting
$N_w$  wires is modelled by equating it to $\exp(iH)$,
where $H$ is a $N_w\times N_w$ Hermitian matrix chosen as follows:
each element along or above the main diagonal is a real number chosen at
random and independently of the other elements of $H$ between $-\pi$ and $\pi$,
and the elements below the diagonal are chosen to ensure $H=H^T$.  For a
superconducting sample the same procedure is adopted, except that the
matrices are of size $2N_w\times 2N_w$ and the elements of $H$ are restricted
to satisfy particle-hole symmetry.  To obtain the scattering matrix for the
whole network, we employ
 a numerical $S$-matrix reduction
algorithm~\cite{jeffery:lattice89,jeffery:lattice90},
 details of which are
 explained more fully
elsewhere~\cite{marksimon:flucs94}.

Figure 2 shows the logarithm of the conductance  $G_{13,24}$  as a function
of the number of slices $M$ in the wire. For each value of $M$,
the conductances arising from 100  different realisations of
the disorder are shown as dots in the figure. The inset shows corresponding
results for the logarithm of the transmission coefficient $T_{13}$. Clearly
the typical values of both
$G_{13,24}$ and $T_{13}$ decay exponentially with $M$ for large $M$. The
different localization lengths for the two systems are to be expected,
since the requirement that particle-hole symmetry must be satisfied for
the superconductor will influence the level statistics of the random node
scattering matrices.
For a superconducting system,
figure 3 shows corresponding results for $G_{13,24}$ (plotted
on a linear scale) and $T_{13}$. This shows that for large $M$, whereas the
transmission coefficient decays exponentially to zero,
the conductance remains finite and can have arbitrary sign.
This confirms our expectation in the strongly localized limit,
 based on equation (6) and additionally shows that
 negative conductances arise in the
presence of quasi-particle transmission.

In the above simulations
positive and negative conductances occur with
roughly equal frequency,  because our choice
of random node scattering matrices  favours neither normal nor Andreev
scattering.
 Historically, the
phenomenon of Andreev scattering, which yields charge transport in the
absence of quasi-particle transmission, was used to explain the marked
difference between thermal and electrical conductance across
normal-superconducting interfaces. In the absence of quasi-particle
transmission, the ends of the sample scatter quasi-particles
independently and therefore apart from a dependence of the condensate potential
on the applied potential
difference~\cite{here:2probe91,here:2probe93,here:multi93},
the voltage probes become completely
decoupled.  In this limit, the voltage applied to probe 1 (3) serves to
cancel the current due to quasi-particles from lead 2 (4) and therefore a
4-probe measurement can be viewed as two independent measurements of
quasi-particle charge imbalance at the two ends of the system.
In the absence of inelastic scattering there is no difference in principle
between charge imbalance measurements~\cite{tinkham} and point contact
spectroscopy
of the kind described by Tsoi and Yakovlev~\cite{Tsoi1}. Here the
sign of the voltage due to quasi-particles was shown to be reversed
by the application of a magnetic field. By analogy such experiments,
one expects
negative 4-probe conductances of the kind predicted in this letter,
 to be particularly
sensitive to the presence of applied or internal magnetic fields.

This work has benefitted from useful correspondance with Marcus B\"uttiker,
who drew our attention to the results of ref.~\cite{buttiker:rev92}.
Financial support in the UK
from the Science and
Engineering Research Council and the Ministry of Defence is gratefully
acknowledged, and one of us (MJ) would like to thank the Science Foundations of
the United States and Japan for a fellowship for research in Japan.

\pagebreak

\pagebreak
\centerline{\bf Figure Captions}

1.(a) A  mesoscopic structure connected to four external probes
numbered 1,2,3,4.
1.(b) A network of one dimensional normal wires connected at nodes,
 with 3 wires in
each external lead. The scatterer consists of
$M$ slices of nodes. For a normal system, the scattering matrix of each node
scatters particles into particles. For a superconducting
system, the scattering matrix of each node also
incorporates Andreev scattering.

2. The main figure is a
scatter plot of the conductances $G_{13,24}$
 of a normal system as a function of
the number of slices $M$. The inset is a corresponding plot of
the transmission coefficients
$T_{13}$. All quantities are plotted on a logarithmic scale.

3. Results for the conductances and transmission coefficients of
a superconducting system. In this case since the conductance no longer
decays exponentially with $M$, only the transmission coefficients are plotted
on a logarithmic scale.


\begin{thebibliography}{10}
\bibitem{note} Strictly, we have only
shown that the conductance may become comparable to $\exp 2L/\xi$,
since we have neglected terms of order $-\exp 2L/\xi$
\end{thebibliography}
\end{document}